\documentclass[12pt]{iopart}

\usepackage{epsfig}
\begin{document}

\title[Wavelength and intensity dependence of multiple forward scattering]{Wavelength and intensity dependence of multiple forward scattering of electrons at above-threshold ionization in mid-infrared strong laser fields}

\author{Chengpu Liu and Karen Z. Hatsagortsyan}

\address{Max-Planck-Institut f\"ur Kernphysik\\ Saupfercheckweg 1, D-69117 Heidelberg, Germany}

\ead{k.hatsagortsyan@mpi-k.de}

\begin{abstract}
The contribution of multiple forward scattering in Coulomb focusing 
of low-energy photoelectrons at above-threshold ionization in mid-infrared laser fields is investigated. 
It is shown that the high-order forward scattering can have a nonperturbative effect in Coulomb focusing. The effective number of rescattering events is defined and is shown to depend weakly on laser intensity and wavelength. Nevertheless, the relative contribution of forward
scattering in Coulomb focusing and the Coulomb focusing in
total decrease with increasing laser intensity and
wavelength.

\end{abstract}



\section{Introduction}

The Coulomb field of the atomic core can play a significant role
in the strong-field photoionization process essentially modifying
the dynamics of low-energy electrons.  It is responsible, in
particular, for the appearance of a rich
structure in the momentum distribution of photoelectrons near the
ionization threshold
\cite{Ullrich,Rudenko,Cock,Faisal,Burgerdorfer,CDLin,RussiaLP09,Burenkov},
for frustrating the tunneling ionization \cite{Eichmann} and for
the creation of a low-energy structure (LES) in
photoelectron spectra in mid-infrared laser fields
\cite{DiMauroNP09,XuPRL09,FaisalNP09,LiuPRL2010,Bauer}.
The Coulomb field focuses low-energy electrons
towards the laser polarization direction which is
mostly due to multiple rescattering \cite{CorkumPRL93} of ionized
electrons by the atomic core at large impact parameters and is termed
Coulomb focusing (CF)
\cite{BrabecIvanov,YudinIvanov,Villeneuve}. For a theoretical
description of Coulomb field effects, different modifications of
the strong field approximation \cite{SFA} have been developed
\cite{Faisal,CC,CC_Lin,CC_Yudin,CC_Popruzhenko}. In addition, the classical
trajectory Monte Carlo (CTMC) method has been successfully employed,
see e.g. \cite{BambiHu97,BurgPRA04,Villeneuve,RussiaLP09,XuPRL09,LiuPRL2010},
for estimation of the effects which are not intrinsically quantum
mechanical.

Recently, the strong field physics in  mid-infrared laser fields
has attracted a lot of attention in connection with the possibility of improving
high-order harmonic generation with mid-infrared driver fields
\cite{DiMauro_HHG}. In mid-infrared laser fields, when 
the Keldysh parameter is small $\gamma=
\sqrt{I_p/2U_p}\ll 1$, the electron dynamics after tunneling 
is mainly classical. This is because the characteristic energies of
the process, $I_p$ and $U_p$, greatly exceed the photon energy in
this regime $\omega\ll I_p\ll U_p$. Here, $I_p$ is the ionization
potential, $U_p=E_0^2/4\omega^2$ the ponderomotive energy, $E_0$
and $\omega$ are the laser field amplitude and frequency,
respectively (atomic units are used throughout). In this regime, the classical features of the three-step
model \cite{CorkumPRL93} are conspicuous and not obscured by interference
effects. Two recent experiments by Blaga et al. \cite{DiMauroNP09}
and Quan et al. \cite{XuPRL09} on the photoionization of atoms and
molecules in strong mid-infrared laser fields reveal a
characteristic spike-like LES in the energy distribution of
electrons emitted along the laser polarization direction. The 
CF is responsible for the effect
\cite{FaisalNP09,LiuPRL2010,Bauer}. More concretely, the LES arises due to
multiple forward scattering (FS) by Coulomb field \cite{LiuPRL2010}.
The CF is usually predicted to decrease with increase in the laser
intensity and wavelength because the average rescattering velocity
and the impact parameter increase in such circumstances. As a consequence,
one may expect that the contribution of high-order FS should also
decrease.
In this context, it was surprising that at a large wavelength of
the mid-infrared laser field, the multiple FS plays a decisive
role for the creation of the LES.

In this paper, we investigate how the contribution of different components of CF  depends on laser intensity and wavelength. Our investigation is limited to the classical interaction regime in mid-infrared laser fields.
Separate components of CF are identified which scale differently with laser parameters: CF which happens
immediately after ionization - initial CF (ICF); CF due to the electron
FS on recollision with atomic core, and
asymptotic CF (ACF) when the electron momentum is disturbed by the
Coulomb field after the laser pulse is switched off. Special attention is devoted to the contribution of the high-order FS events and to the definition of the effective number of FS
events. We use the
CTMC method with tunneling and the Coulomb field of the atomic
core fully taken into account.

\section{The method}
\label{method}

The 3D CTMC method employed in this paper is developed as follows.
(1) An ensemble of electrons is formed corresponding to the
tunneled electron wave packet according to the
Ammosov-Delone-Krainov (ADK) theory \cite{ADK}. The electrons are born
with the following initial conditions. The electron initial position along the laser polarization
direction is derived from the Landau's effective potential theory
\cite{Landau77}. The transverse coordinates of the initial position are zero. The initial longitudinal momentum is zero   and the  transverse one follows the corresponding ADK distribution \cite{ADK2}.
(2) The  electron wave-packet propagates in
the field of a laser pulse and Coulomb potential via the
solution of Newton equations. (3) The positions and momenta of
electrons when the laser pulse is switched off are used to calculate the
asymptotic momenta at the detector \cite{Landau_Mechanics}. (4)
Each trajectory  is weighted by the ADK ionization rate
and the initial transverse momentum distribution function
\cite{ADK2}. (5)
The shape of the laser pulse is half-trapezoidal: For the first ten cycles, the field has
a constant amplitude and is ramped off within the last three cycles.
The electrons are launched within the first half cycle
since there are no multi-cycle interference effects in the classical calculation.
The ensemble consists of $ 10^6 $ particles  and the convergence is
checked via double increase of this number. The target atom is neon
with ionization potential $I_p = 21.56$ eV which can endure a
maximum laser intensity $I_0\approx 8.66\times 10^{14}$ W/cm$^2$ \cite{criticalintensity}.
The process is in the tunneling regime, e.g., $\gamma\approx 0.2 $ at $I_0 =
7.24 \times 10^{14}$ W/cm$^2$ and wavelength $\lambda = 2 \mu $m.

We estimate the contributions of the multiple FS, ICF
 and ACF to the total CF in the following way:
(1) For each rescattering event at the moment $t_s$, the minimal
distance from the core $r_s$,  the distance from the core in the
transverse plane (with respect to the laser polarization
direction) $\rho_s$ and the electron momentum $p_s$ are determined
numerically. Then, the transverse momentum change $\delta p_{\bot}$ due to the Coulomb
potential $V(r)$ at the $s$-th forward scattering event is
estimated as $\delta p_{\bot \,s} \approx \int
\nabla_{\bot}V(r(t))dt\sim -(\rho_s/r_s^3) \delta t_s$, where
$\delta t_s$ is the rescattering time duration. When the electron
velocity $p_s$ in the FS event is large, $\delta t_s\sim 2r_s/p_s$. In the
opposite case, $\delta t_s\sim 2 \sqrt{2r_s/|E(t_s)|}$ is
determined by the laser field $E(t_s)$ at the $s$-th rescattering moment $t_s$.
Accordingly,
\begin{eqnarray}
\delta p_{\bot \,s} &=& -{2\rho_s}/{(r_s^2 p_s)}, \quad {\rm
if}\quad p_s^{2} \gg r_s |E(t_s)|\,
\nonumber \\
\delta p_{\bot \,s} &=& -{2 ^{3/2} \rho_s}/{
\sqrt{r_s^5|E(t_s)|}}, \quad {\rm otherwise}. \label{FS}
\end{eqnarray}
(2) The transverse momentum change due to ICF is estimated
numerically as the deviation of the exact transverse momentum from that neglecting the Coulomb potential, 
after a half laser period following the ionization
moment $t_i$: $\delta p_{\bot}^{(I)}=p_{\bot}
(t_i+T/2)-p_{\bot}^{(NC)} (t_i+T/2)$, where $T$ is the laser
period and $p_{\bot}^{(NC)}$ is the electron transverse momentum
neglecting the Coulomb field. The numerical estimate for $\delta
p_{\bot}^{(I)}$ is slightly larger in absolute value than the
analytical one \cite{RussiaLP09}:
\begin{eqnarray}
\delta p_{\bot}^{(I)}
\approx -2p_{i\bot}|E(t_i)|/(2I_p)^2,
\label{ICF}
 \end{eqnarray}
with the initial transverse momentum $p_{i\bot}$. (3) We estimate the ACF contribution via numerical
comparison of the asymptotic electron momentum with the one after
switching off the laser pulse.

\section{The results}

\begin{figure}
\begin{center}
\includegraphics[width=0.75\textwidth,angle=0]{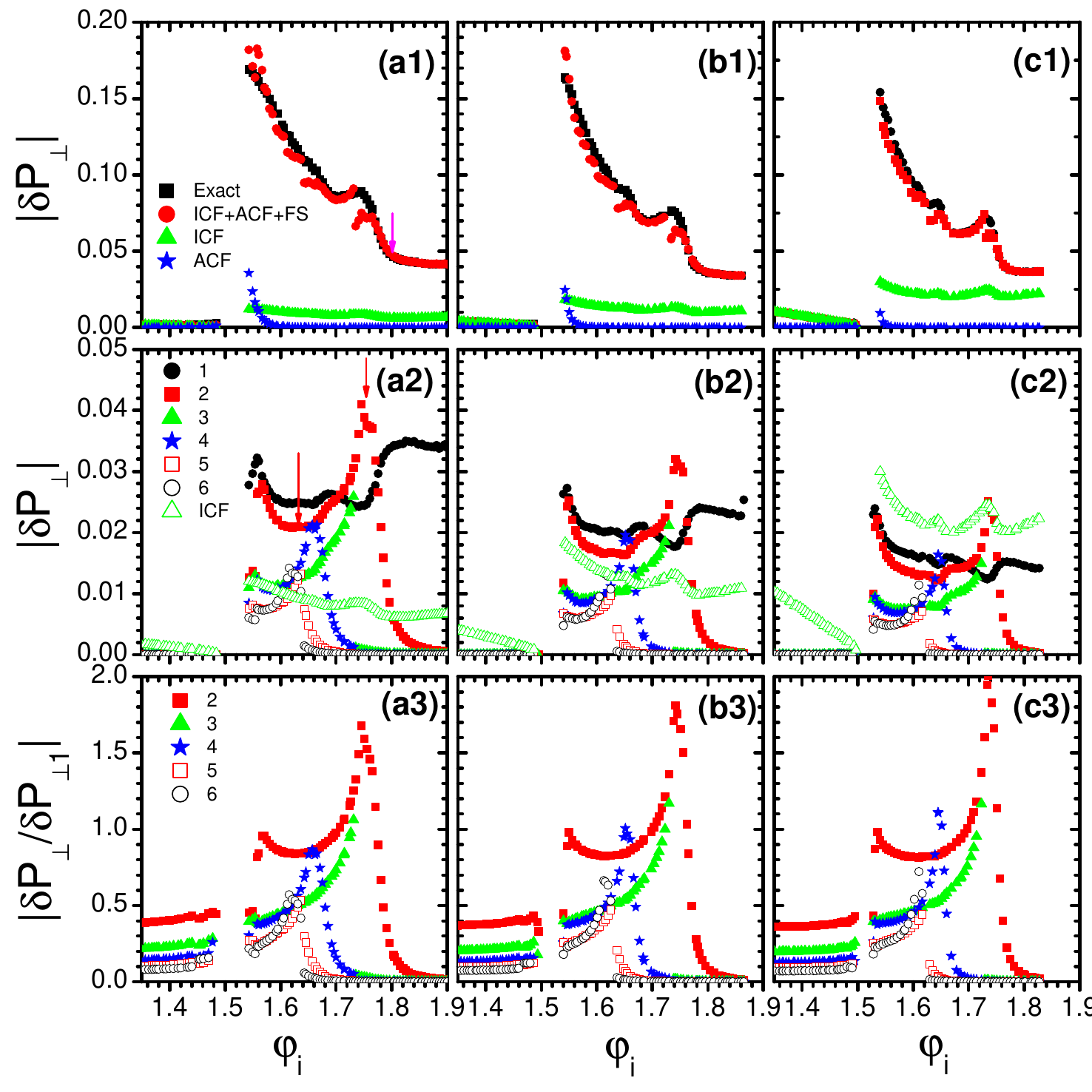}
\caption{(color online) The  transverse momentum change $\delta p_{\bot}$ versus the ionization phase $\varphi_i = \omega t_i$. The CTMC simulation for a neon atom in
a mid-infrared laser field with a wavelength $\lambda = 2 \mu m$ for
the following laser intensities: (a1-a3) $I=1.81\times 10^{14}$ W/cm$^2$, (b1-b3) $I=3.62\times 10^{14}$ W/cm$^2$  and  (c1-c3) $I=7.24\times 10^{14}$ W/cm$^2$. (a1,b1,c1) The
total transverse momentum change (marked as ``exact''), the estimation of ICF
and ACF as well as of the total transverse momentum change (marked as ``ICF+ACF+FS'') as described in Sec.\ref{method}. (a2,b2,c2)  $\delta p_{\bot}$ due to the $s$-th order FS events ($s$ is indicated in the inset) and due to ICF. (a3,b3,c3) The ratio of the $\delta p_{\bot}$ at the $s$-th order FS events  to  the first-order one. The ionization phase $\varphi_{i}^{(1)}$ corresponding to the threshold of the multiple FS is marked by an arrow in (a1). The peak and the plateau of $\delta p_{\bot}$ for the 2nd FS are marked by arrows in (a2). The maximum of the laser field is at $\varphi_i=\pi/2$. 
} \label{Intensity}
\end{center}
\end{figure}

\begin{figure}
\begin{center}
\includegraphics[width=0.75\textwidth,angle=0]{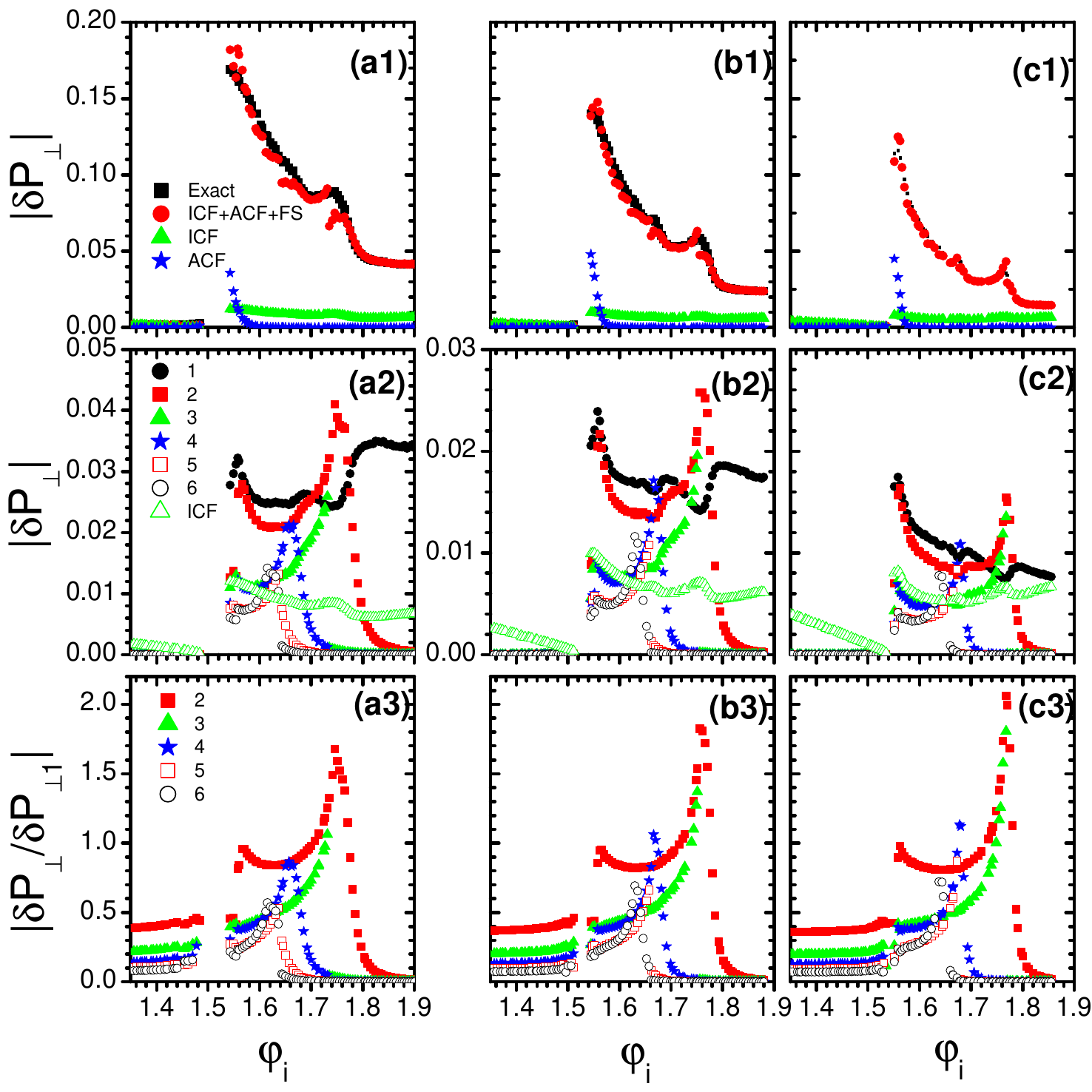}
\caption{(color online) The  transverse momentum change $\delta p_{\bot}$ versus the ionization phase $\varphi_i$. The CTMC simulation for a neon atom in a
mid-infrared laser field with an intensity of $I=1.81\times 10^{14}W/cm^2$ for
the following wavelengths: (a1-a3)  $\lambda = 2 \,\mu $m, (b1-b3)  $\lambda = 2.828 \,\mu $m and
(c1-c3) $\lambda = 4 \,\mu $m.  (a1,b1,c1) The
total transverse momentum change (marked as ``exact''), the estimation of ICF
and ACF as well as of the total transverse momentum change (marked as ``ICF+ACF+FS'') as described in Sec.\ref{method}. (a2,b2,c2)  $\delta p_{\bot}$ due to the $s$-th order FS events ($s$ is indicated in the inset) and due to ICF. (a3,b3,c3) The ratio of the $\delta p_{\bot}$ at the $s$-th order FS events  to  the first-order one. 
} 
\label{Wavelength}
\end{center}
\end{figure}

The CF is mainly due to multiple small-angle scattering. It is significant only for low energy photoelectrons, and we will examine the dynamics for such electrons in details. The CF is characterized by the transverse momentum change $\delta p_{\bot}$
induced by the Coulomb field \cite{LiuPRL2010} which depends on the ionization phase $\varphi_i\equiv \omega t_i$. We restrict ourselves to ionization phases and to trajectories which contribute to the low energy part (up to $40$ eV) of above-threshold ionization spectrum emitted along the laser polarization direction within an opening angle of $\pm 2.5^{\rm o}$. The electrons, which are emitted out of the laser polarization direction, have experienced large-angle scattering, their CF is interrupted and, consequently, their dynamics is not typical for CF. The laser intensity dependence of the different CF components is shown in Fig. \ref{Intensity} and the wavelength dependence in Fig. \ref{Wavelength}. For each $\varphi_i$, the transverse momentum change is shown for the electron trajectory which has  the maximal probability among the contributing trajectories at this ionization phase. We calculate the total transverse momentum change exactly via the CTMC simulation, see the curves marked as ``exact'' in Figs.  \ref{Intensity}
(a1,b1,c1) and \ref{Wavelength} (a1,b1,c1). Further in Figs. \ref{Intensity} and \ref{Wavelength}, we show the results of the estimate of $\delta p_{\bot}$ due to the s-th FS ($s\leq 6$), ICF and ACF as described in Sec.\ref{method}.
To show the accuracy of our estimations, we sum up all contributions to $\delta p_{\bot}$ and comapare it with the exact result, see the curves  marked as  ``ICF+ACF+FS'' in Figs.  \ref{Intensity} (a1,b1,c1) and \ref{Wavelength} (a1,b1,c1).

From the analysis of Figs. \ref{Intensity} and \ref{Wavelength}, the following conclusions can be drawn. First of all, Figs. \ref{Intensity}
(a1,b1,c1) and \ref{Wavelength} (a1,b1,c1) show that the contribution of
ACF to the total $\delta p_{\bot}$ is generally
negligible (the main contribution is at ionization phases near the
peak of the laser field within the ionization phase interval of $\delta\varphi_i\approx 0.02$). It decreases with increasing intensity and  does not change with  wavelength.
The contribution of ICF  to the total $\delta p_{\bot}$ increases with  increasing  intensity and
remains almost constant with  wavelength which is consistent with the estimate of Eq.~(\ref{ICF}). 
The contribution of ICF still constitutes a small fraction of the total $\delta
p_{\bot}$ (less than 10\%) for ionization phases
$\pi/2<\varphi_i<\varphi_{i}^{(1)}$ (the maximum of the laser field is at $\varphi_i=\pi/2$), where multiple scattering takes place but competes with the single scattering contribution
at $\varphi_i>\varphi_{i}^{(1)}$, especially at high intensities
and wavelengths (see the estimate below, Eq. (\ref{ICF/FS1})). The
ionization phase $\varphi_{i}^{(1)}$ marks the threshold of the
multiple FS, see the indication of $\varphi_{i}^{(1)}$ in Fig. \ref{Intensity} (a1); $\varphi_{i}^{(1)} \approx 1.8$ at a laser intensity of $1.8\times 10^{14}$ W/cm$^2$ but decreases
slightly with  increasing intensity.

\begin{figure}
\begin{center}
\includegraphics[width=0.45\textwidth,angle=0]{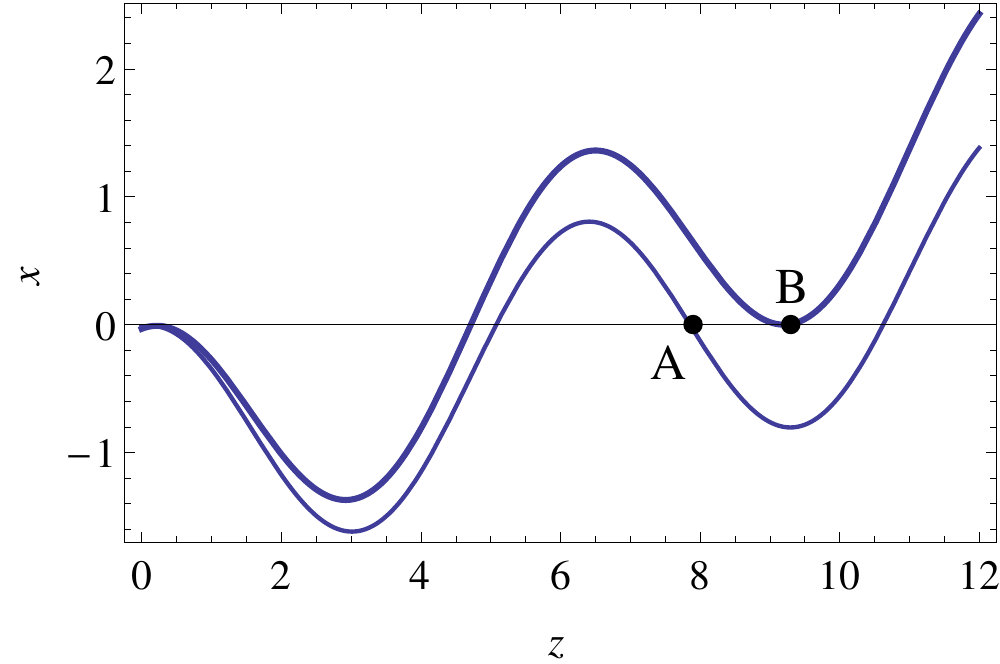}
\caption{  The electron trajectories at $\varphi_i=1.75$ (thick line) and $\varphi_i=1.64$ (thin line). The $x$-coordinate along the laser polarization direction is scaled by $E_0/\omega^2$ and the transverse $z$-coordinate by $p_{i\bot}/\omega$. The position A corresponds to the  plateau  and  B to the peak of the transverse momentum change at the second forward scattering.
} \label{trajectories}
\end{center}
\end{figure}

The main CF contribution to the total $\delta p_{\bot}$ comes from FS which determines the
shape of the curve $\delta p_{\bot}$ versus the ionization phase, shown in Figs. \ref{Intensity}
(a1,b1,c1) and \ref{Wavelength} (a1,b1,c1). 
The transverse momentum change $\delta p_{\bot \,s}$ due to the $s$-th order FS in the case of different laser intensities and wavelengths
are shown in Figs. \ref{Intensity} (a2,b2,c2) and \ref{Wavelength}
(a2,b2,c2), respectively. $\delta p_{\bot \,s}$ has a characteristic
dependence on the electron ionization phase which is qualitatively
the same for each scattering order. The $\delta p_{\bot \,s}$ increases sharply with decrease in the ionization phase from the threshold value (it is different for different scattering orders), reaches the peak and then decreases
slowly  down  to a flat plateau, the latter having an increasing tail
with the further decrease of the ionization phase. Although the contribution of FS decreases
on average with increasing order, a higher-order FS can make a larger
contribution in some phase intervals than a lower-order one.

\begin{figure}
\begin{center}
\includegraphics[width=0.75\textwidth,angle=0]{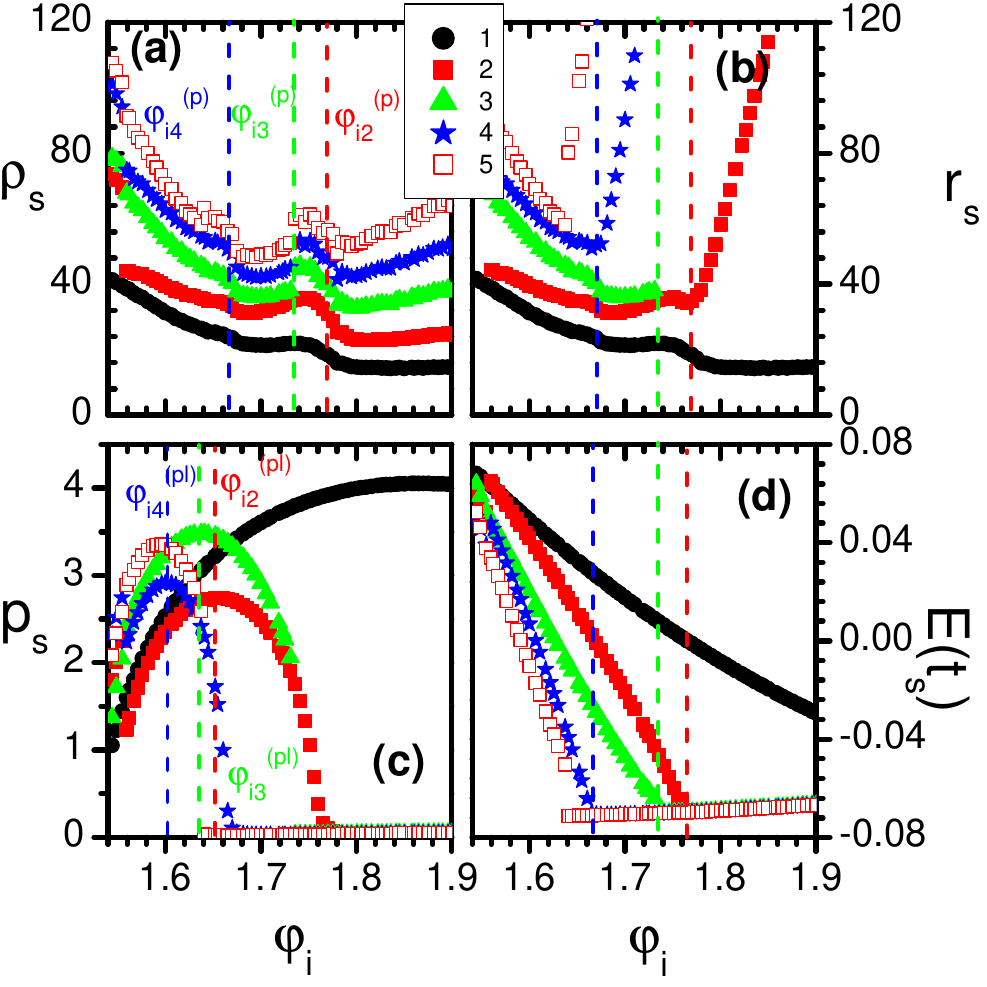}
\caption{(color online) The parameters of the $s$-th order FS ($s$ is indicated in the inset): (a) $\rho_s$ the distance from the atomic core in the transverse plane, (b) $r_s$ the distance from the core, (c) $p_s$ the momentum, (d) $E(t_s)$ the field at the scattering moment.   $\varphi_{i\,s}^{(p)}$ and $\varphi_{i\,s}^{(pl)}$ are the ionization phases corresponding to the peak and plateau for the $s$-th order FS, respectively, which are indicated with dashed lines. The CTMC simulation for a neon atom in a laser field with $I=1.81\times 10^{14}W/cm^2$ and  $\lambda = 2 \,\mu $m.  
} \label{parameters}
\end{center}
\end{figure}

\begin{figure}
\begin{center}
\includegraphics[width=0.75\textwidth,angle=0]{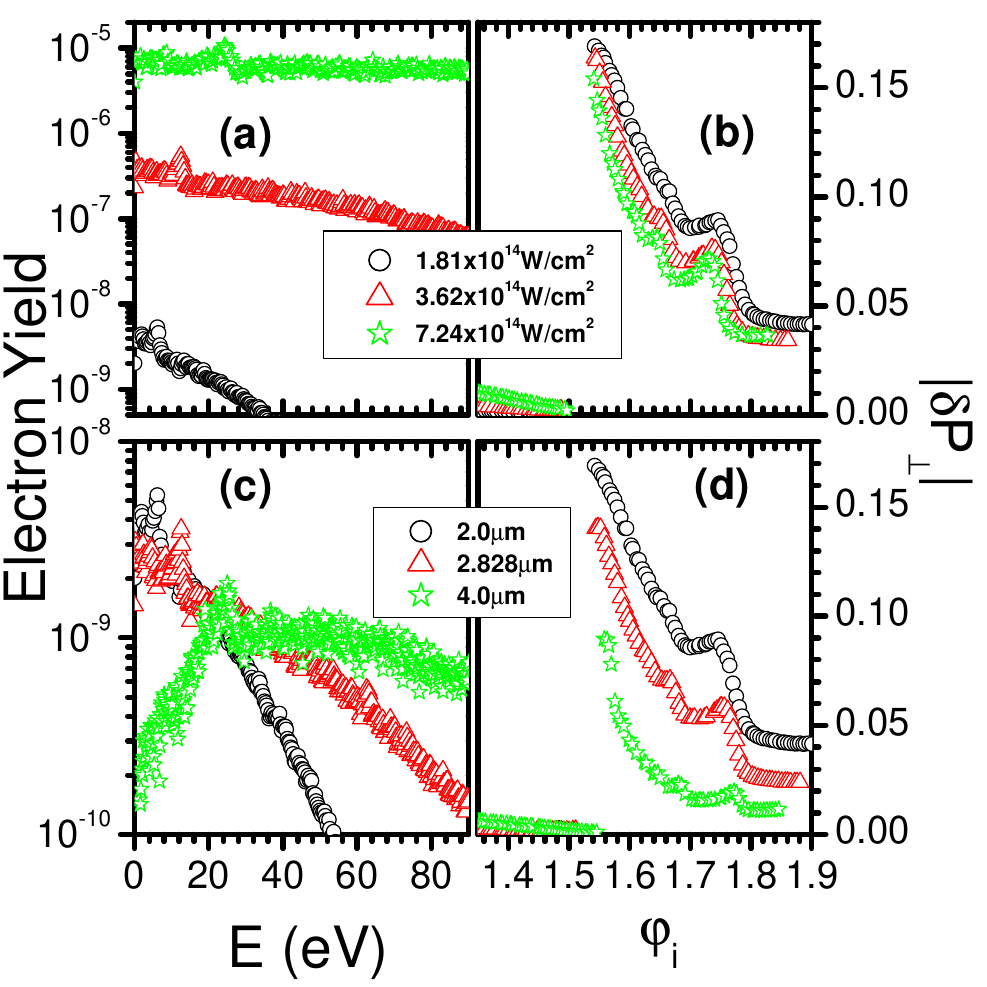}
\caption{(color online) The CTMC simulation for a neon atom in
a mid-infrared laser field: (a,c) Photoelectron spectra. (b,d) The transverse momentum change  versus $\varphi_i$. (a,b) $\lambda = 2 \mu m$ for
different laser intensities indicated in the inset. (c,d)  $I=1.81\times 10^{14}W/cm^2$  for
different wavelengths indicated in the inset.  
} \label{spectra}
\end{center}
\end{figure}

Let us estimate the
values of the peaks and plateaux of the transverse momentum
change due to high-order FS. The peak in the $\delta p_{\bot
\,s}$ for the $s$-th order FS  arises when the electron trajectory touches
the $z$-axis at a recollision (the coordinate center is chosen at the atomic
center, $x$-axis is in the laser polarization direction and
$z$-axis in the transverse direction, see Fig. \ref{trajectories}). In this case,
$r_s\approx\rho_s$, $E(t_s)\approx E_0$, the electron
momentum $p_s$ is nearly zero, see Fig. \ref{parameters} (a-d), and
$\delta p_{\bot \,s}$ is determined by the second expression of
Eq. (\ref{FS}). The ionization phases corresponding to the peak and the plateau of the s-th order FS are indicated in Fig. \ref{parameters} by $\varphi_{is}^{(p)}$ and $\varphi_{is}^{(pl)}$, respectively. The impact parameter at the peak of the $s$-th order FS
can be estimated as $\rho_s^{\rm peak}\sim p_{i\bot}\pi(s+1)/\omega$, see Fig. \ref{trajectories}.
Then, using Eq.(\ref{FS}) we have
\begin{eqnarray}
|\delta p_{\bot \,s}^{\rm peak}| \sim
\frac{1}{\sqrt{E_0}}\left(\frac{2\omega}{\pi
p_{i\bot}(s+1)}\right)^{3/2}. \label{peak}
\end{eqnarray}
The plateau in the $\delta p_{\bot \,s}$ corresponds to the FS
case when the electron velocity in the FS point is the largest, see Fig. \ref{parameters} (c).
The latter is determined by the amplitude of the velocity
oscillation in the laser field: $v_s\approx \beta_s E_0/\omega$, with $\beta_s\approx
0.8$ at even $s$ and $\beta_s\approx 1$ at odd $s$, according to the numerical results. The impact
parameter in this case is estimated as $\rho_s^{\rm plateau}\sim p_{i\bot}\pi
s/\omega$, which yields
\begin{eqnarray}
|\delta p_{\bot \,s}^{\rm plateau}| \sim   \frac{2\omega^2}{\pi
p_{i\bot}E_0\beta_s s} . \label{plateau}
\end{eqnarray}
The estimates of Eqs. (\ref{peak}) and (\ref{plateau}), which are in
agreement with the numerical calculations presented in Figs.
\ref{Intensity} and \ref{Wavelength}, show that the peaks for the higher
order FS ($(s+1)$-th order) can exceed the plateaux of the lower-order FS ($s$-th order). In fact, this ratio is 
\begin{eqnarray}
|\delta p_{\bot \,s+1}^{\rm peak}|/|\delta p_{\bot \,s}^{\rm
plateau}| \approx \sqrt{\frac{2E_0}{\pi p_{i
\bot}\omega}}\frac{s\beta_s}{(s+2)^{3/2}},
 \label{ratio}
\end{eqnarray}
which is between $1.1$ and $1.2$ for $s=2-6$ at $1.81\times 10^{14}$
W/cm$^2$. Especially, the peaks of the even order FS  (2nd, 4th,...) are larger than
the corresponding odd FS plateaux   (1st, 3rd,...). The plateaux of the  even order  (2nd, 4th,...) 
FS are comparable with that of the corresponding odd FS (1st, 3rd,...). These are because the velocity at even FS events is smaller
than that at odd FS. Eq. (\ref{ratio}) shows a remarkable feature  that the
peak-to-plateau ratio increases with increasing intensity and wavelength.
In particular, due to a larger contribution of the 6th order FS peak with respect to the plateau of the 5th order FS at intensity $7.24\times 10^{14}$ W/cm$^2$, see Fig. \ref{Intensity} (c2), an
additional oscillation in the $\delta p_{\bot}$ dependence on the
ionization phase arises, at $\varphi_i\approx 1.64$ in
Fig.\ref{spectra} (b), which  induces an additional lower energy
peak (at about $9$ eV) in the energy distribution within the LES, see Fig.
\ref{spectra} (a). 

Using Eqs. (\ref{ICF}) and (\ref{plateau}), we calculate 
the ratio of ICF to the plateau of the first-order FS:
\begin{equation}\label{ICF/FS1}
\frac{\delta p_{\bot}^{(I)}}{\delta p_{\bot \,1}^{\rm
plateau}}=\pi\left(\frac{
p_{i\bot}}{2I_p}\frac{E_0}{\omega}\right)^2.
\end{equation}
The latter confirms the above statement that ICF can compete
with the single scattering contribution to CF at high intensities
and wavelengths. For instance, at typical parameters $p_{i\bot}=0.1$ a.u., $I_p=0.79$ a.u., $I=7\times 10^{14}W/cm^2$ and $\lambda = 2 \mu m$ the ratio  in Eq. (\ref{ICF/FS1}) amounts to $\delta p_{\bot}^{(I)}/\delta p_{\bot \,1}^{\rm plateau}\approx 0.4$.

The contribution of the FS (as well as the total CF effect)
decreases with increasing intensity and  wavelength
because of the increased scattering velocity and the impact
parameter, see Eqs. (\ref{peak}) and (\ref{plateau}). For the same reason the curves in the phase space move down with higher intensities and wavelengths in Figs. \ref{spectra} (b) and \ref{spectra} (d).
However, the ratio of $\delta p_{\bot}$ of different scattering orders does not significantly variate with variation of laser intensity and wavelength. Thus, the peak of the transverse momentum change due to the $s$-th FS scaled with that of the first FS can be estimated
\begin{eqnarray}
|\delta p_{\bot \,s}^{\rm peak}|/|\delta p_{\bot \,1}^{\rm
plateau}| \approx \sqrt{\frac{2E_0}{\pi p_{i
\bot}\omega}}\frac{1}{(s+1)^{3/2}}.
 \label{ratio2}
\end{eqnarray}
It shows that the relative role of the s-th FS even incerases slowly with increasing laser intensity and wavelength. This can also
be seen from Figs. \ref{Intensity} (a3,b3,c3) and \ref{Wavelength} (a3,b3,c3). The mentioned feature can be interpreted as a slow variation of the effective number of scattering.  

\section{The effective number of scattering}

The effective number of scattering $N_{eff}$ can be defined
employing Eqs. (\ref{peak}) and (\ref{plateau}). We define $N_{eff}$ as follows. First of all, we choose a small parameter $\epsilon \ll 1$ to determine the accuracy in which the contributions of the high-order FS to the total $\delta p_{\bot}$ can be neglected with respect to that of the first FS $\delta p_{\bot 1}$. The effective number of FS $N_{eff}$ is defined to determine the highest order of FS which makes nonnegligible contribution to the total $\delta p_{\bot}$. Namely, at a given $\epsilon$,  
\begin{eqnarray}
\label{sss}
  \delta p_{\bot \,s}^{\rm peak}/\delta p^{\rm plateau}_{\bot 1}&>&\epsilon, \,\,\,\,\,{\rm if} \,\,s<N_{eff}, \nonumber \\ 
\delta p_{\bot \,s}^{\rm peak}/\delta p^{\rm plateau}_{\bot 1}&<&\epsilon,\,\,\,\,\,{\rm if} \,\,s>N_{eff}.
\end{eqnarray}
The latter definition yields the following expression for the effective number of FS 
\begin{eqnarray}
\label{bar{s}3}
 N_{eff}\approx \frac{1}{\epsilon^{2/3}}\left(\frac{2}{\pi
 }\frac{E_0}{\omega p_{i\bot}}\right)^{1/3}-1.
\end{eqnarray}
For the intensity range shown in Fig. \ref{Intensity}, our
criterion gives for $N_{eff}$ a number between 5 and 6 (for the
concreteness $\epsilon=0.1$ is assumed). The number of scattering according to Eq.(\ref{bar{s}3})
increases slowly with increase in the laser intensity and wavelength.

Generally, the total CF effect decreases with intensity and
wavelength because the main contribution in CF comes from the
multiple forward scattering which, in total, decreases as expected,
see Fig. \ref{spectra} (b) and (d), where the
curves in the phase space move down, generally,  with increasing
laser intensity and wavelength. The exception is the tail at
$\varphi_i>\varphi_{i}^{(1)}$ in Fig. \ref{spectra} (b) which is
due to the competition of the first FS with the initial CF as mentioned above.
The most important property is that the ratio of the transverse
momentum changes due to FS of different orders are almost constant
with increasing  laser intensity and wavelength, see Figs.
\ref{Intensity} (a3,b3,c3) and \ref{Wavelength} (a3,b3,c3). Therefore,
the shape of the phase space distribution (dependence of  $\delta p_{\bot}$ on the ionization phase) remains similar, see Fig. \ref{spectra} (b) and  (d).
Note that the consecutive slope changes of the phase space
distribution are responsible for the creation of LES and are determined by the contributions of the second, third and forth FS \cite{LiuPRL2010}.

 We point out some features which distinguish the wavelength
dependence of CF from the intensity dependence, see Fig.
\ref{Wavelength}.
The FS  contribution to the total $\delta p_{\bot}$  decreases more strongly with
increasing wavelength as Eqs. (\ref{peak}) and (\ref{plateau}) indicate.
Due to the latter, the knee of the phase space distribution becomes less
prominent at higher wavelengths. The relative contribution of ICF to the total $\delta p_{\bot}$ with respect to FS increases with increasing wavelength but to a lesser extent than in the case of the intensity dependence. This is because of the $p_{i\bot}$ factor in Eq. (\ref{ICF/FS1}) which
decreases with increase in the wavelengths. 

\section{Conclusion}

At above-threshold ionization in the realm of intensities and wavelengths
corresponding to the classical regime $\omega \ll I_p$ and $\gamma
\ll 1$, multiple forward scattering of an ionized electron has
a nonperturbative contribution to Coulomb focusing. In some regions
of ionization phase (photoelectron energy), the contribution of the higher-order forward scattering  to the total Coulomb focusing  can dominate the lower-order one which creates local peaks in the
photoelectron spectra. The effective number of scattering  does not depend significantly on laser intensity and
wavelength. 

\section*{Acknowledgments}

The fruitful discussions with C. H. Keitel are acknowledged.


\section*{References}
\begin{thebibliography}{99}

\bibitem{Ullrich} Moshammer R \emph{et al.} 2003 \emph{Phys. Rev. Lett.} {\bf 91} 113002

\bibitem{Rudenko} Rudenko A \emph{et al.} 2004 \emph{J. Phys. B} {\bf 37} L407

\bibitem{Cock} Maharajan C M 2006 \emph{J. Phys. B} {\bf 39} 1955

\bibitem{Faisal} Faisal F and Schlegel G 2005 \emph{J. Phys. B}  {\bf 38} L223

\bibitem{Burgerdorfer} Arbo D G \emph{et al.} 2006 \emph{Phys. Rev. Lett.} {\bf 96} 143003

\bibitem{CDLin} Wickenhauser M \textit{et al.} 2006 \emph{Phys. Rev. A} \textbf{74} 041402(R)

\bibitem{RussiaLP09} Shvetsov-Shilovski N I \textit{et al.} 2009 \emph{Laser Phys.} {\bf 19} 1550

\bibitem{Burenkov} Burenkov I \textit{et al.} 2010 \emph{Laser Phys. Lett.} \textbf{7} 409

\bibitem{Eichmann}  Nubbemeyer T \emph{et al.} 2008 \emph{Phys. Rev. Lett.} \textbf{101} 233001\\
Eichmann U \emph{et al.} 2009 \emph{Nature} {\bf 461} 1261

\bibitem{DiMauroNP09} Blaga C \textit{et al.} 2009 \emph{Nature Phys.} {\bf 5} 335\\
 Catoire F \emph{et al.} 2009 \emph{Laser Phys.} {\bf 19} 1574

\bibitem{XuPRL09} Quan W \emph{et al.} 2009 \emph{Phys. Rev. Lett.} {\bf 103} 093001

\bibitem{FaisalNP09} Faisal F 2009 \emph{Nature Phys.} {\bf 5} 319

\bibitem{LiuPRL2010} Liu C and Hatsagortsyan K Z 2010 \emph{Phys. Rev. Lett.} {\bf 105} 113003

\bibitem{Bauer} Yan Tian-Min ,  Popruzhenko S V,  Vrakking M J, and  Bauer D 2010
Phys. Rev. Lett. \textbf{105} 253002

\bibitem{CorkumPRL93} Schafer K J \emph{et al.} 1993 \emph{Phys. Rev. Lett.} {\bf 70} 1599\\
Corkum P 1993 \emph{Phys. Rev. Lett.} {\bf 71} 1994

\bibitem{BrabecIvanov} Brabec T, Ivanov M Yu, and Corkum P B 1996 \textit{Phys. Rev. A} \textbf{54} R2551

\bibitem{YudinIvanov} Yudin G L, and Ivanov M Yu 2001 \emph{Phys. Rev. A} {\bf 63} 033404

\bibitem{Villeneuve} Comtois D \emph{et al.} 2005 \emph{J. Phys. B} {\bf 38} 1923

\bibitem{SFA} Keldysh L 1964 \emph{Sov. Phys. JETP} {\bf 20} 1945\\
Faisal F 1973 \emph{J. Phys. B} {\bf 6} L89\\
Reiss H 1980 \emph{Phys. Rev. A} {\bf 22} 1786

\bibitem{CC}  Duchateau G, \textit{et al.} 2001 \emph{Phys. Rev. A} {\bf 63} 053411

\bibitem{CC_Lin} Chen Z \textit{et al.} 2006 \emph{Phys. Rev. A}, \textbf{74} 053405

\bibitem{CC_Yudin} Yudin G L \textit{et al.} 2007 \emph{J. Phys. B}  {\bf 40} F93\\
Yudin G L, Patchkovskii S and Bandrauk A D 2008 \emph{J. Phys. B}  {\bf 41} 045602

\bibitem{CC_Popruzhenko} Popruzhenko S V, Paulus G G, Bauer D 2008 \emph{Phys. Rev. A} \textbf{77} 053409\\
Popruzhenko S V, Bauer D, 2008 \emph{J Mod. Opt.} \textbf{55} 2573

\bibitem{BambiHu97} Hu B, Liu J, and  Chen S G 1997 \emph{Phys. Lett. A} {\bf 236} 533

\bibitem{BurgPRA04} Dimitriou K I,\textit{et al.} 2004 \emph{Phys. Rev. A} {\bf 70} 061401(R)

\bibitem{DiMauro_HHG} Colosimo P \textit{et al.} 2008  \emph{Nature Phys.} \textbf{4} 386  

\bibitem{ADK} Perelomov A M, Popov V S and Teren'ev V M 1967 \emph{Zh. Eksp. Teor. Fiz.} {\bf 52} 514 [1967 \emph{Sov. Phys. JETP }{\bf 25} 336]\\
 Ammosov M V, Delone N B and Krainov V P 1986 \emph{ibid}. {\bf 91} 2008 [1986 \emph{ibid}. {\bf 64}
 1191]

\bibitem{Landau77} Landau L D and Lifshitz E M 1977 \emph{Quantum Mechanics} (Pergamon, Oxford) p. 293

\bibitem{ADK2} Delone N B and Krainov V P 1991 \emph{J. Opt. Soc. Am. B} {\bf 8} 1207

\bibitem{Landau_Mechanics} Landau L D and Lifshitz E M 1993 \emph{Mechanics} (Pergamon, Oxford) p. 170

\bibitem{criticalintensity}
Bethe H and Salpeter E 1977 \emph{Quantum Mechanics of Atoms with One and Two
Electrons} (Berlin: Springer)


\end{thebibliography}
\end{document}